\documentclass[referee]{raa}            

\usepackage{graphicx,times}             
\usepackage{natbib}
\usepackage{amssymb,amsmath}
\usepackage{physics}
\bibpunct{(}{)}{;}{a}{}{,}

\usepackage[pagebackref=true]{hyperref}

\begin{document}

  \title{Pulsar anti-glitches: starquakes driven by magnetism?
}

   \volnopage{Vol.0 (20xx) No.0, 000--000}      
   \setcounter{page}{1}          

   \author{Hanyuan Long
        \inst{1}
     \and Ruipeng Lu
        \inst{2}
     \and Weiyang Wang
        \inst{3}
     \and Shunshun Cao
        \inst{1}
     \and Hao Tong 
        \inst{4}
     \and Han Yue 
        \inst{2}
     \and Renxin Xu
        \inst{1}
}
    \institute{School of Physics, Peking University,
             Beijing 100871, China;  {\tt r.x.xu@pku.edu.cn}\\
             \and
             School of Earth and Space Sciences, Peking University, Beijing 100871, China; {\tt yue.han@pku.edu.cn}\\
             \and
             School of Astronomy and Space Sciences, University of Chinese Academy of Sciences, Beijing 100049, China;  {\tt wywang@ucas.ac.cn}
             \\
             \and
             School of Physics and Materials Science, Guangzhou University, Guangzhou 510006, China\\
\vs\no
   {\small Received 20xx month day; accepted 20xx month day}}

\abstract{In the conventional starquake model of pulsar glitches, it is usually assumed that such events arise from fault slip induced by the self-gravity of compact objects. This inevitably decreases the moment of inertia, producing a glitch with an amplitude of only $\Delta\nu/\nu > 0$.
However, an increasing number of anti-glitches ($\Delta\nu/\nu < 0$) have been observed in extremely magnetized pulsars, the magnetars, and this cannot be explained by that framework.
In the present study, we hypothesis that magnetic stresses within a compact object can make for elastic deformations that trigger fault slipping, resulting in a ``magnetism-driven starquake'' when the local breaking threshold is exceeded.
This process can then either decrease or increase the moment of inertia, naturally generating a glitch or an anti-glitch, respectively.
With an order-of-magnitude calculation in this brief report, we present a simple relationship between the magnetic field $B$ and the amplitude $\Delta\nu/\nu$, which is consistent with the observational distribution of existing glitch and anti-glitch data.
Further discoveries of glitch/anti-glitch events, alongside more quantitative models of elastic–magnetic stress coupling, would be welcome and could eventually provide clear tests for the hypothesis.
\keywords{anti-glitch; magnetar; starquake; magnetic stress}
}

   \authorrunning{Long et al.}
   \titlerunning{Anti-glitches: Magnetism-driven Starquakes?}  

   \maketitle

%
%
\section{Introduction}           
\label{sect:intro}

It is well known that atomic nuclei are microscopic objects of nucleons condensed by the fundamental strong force, with baryon numbers only ranging from $A\sim 10^0$ to $10^2$.
This type of condensed matter can be simply termed ``{\em strong matter}'', in contrast to ordinary matter, which is made up of atoms held together by electromagnetic forces and is likewise known as ``{\em electric matter}''~\citep{2018SCPMA..6109531X}.
Could any other form of strong matter exist besides atomic nuclei?
Yes, it is of the so-called pulsar-like compact objects with $A\sim 10^{57}$!
However, the nature of these kinds of objects remains one of the most fundamental open problems in modern physics, to be relevant to the non-perturbative nature of quantum chromo-dynamics (QCD), which makes a first-principle calculation of the equation of state unrealistic.
Nevertheless, strangeness matters in determining the state of supra-nuclear matter, and it has been argued that pulsar-like objects are actually solid strangeon stars rather than conventional neutron stars.~\citep{2003ApJ...596L..59X,2025IJMPA..4050180X}.
Although strangeon nuggets with $A\sim 10^{30}$ could potentially be detected using acoustic arrays~\citep{2025RAA....25i5010Q}, this presentation focuses on strangeon stars with $A\sim 10^{57}$, where quakes could naturally have a significant impact on observations of these globally solid objects.
We are investigating starquakes driven by magnetism in a strangeon star (or a conventional neutron star). In particular, we are interested in the possibility that a massive strangeon star with a large magnetic Reynolds number may result in an ultra-strong magnetic field (a so-called strangeon magnetar) through dynamo actions~\citep{Xu2024}.

To be specific, we examine whether magnetism-driven starquakes, as an extension of conventional gravity-driven starquake models, can account for a variety of anomalous pulsar timing phenomena, particularly glitches and the recently observed anti-glitches in magnetars.
It is well known that some of the timing irregularity of spin evolution of pulsars can be interpreted by sudden increases in their rotational frequency, known as glitches~\citep{1974IAUS...53..227B}.
Since their first discovery and confirmation in the Vela pulsar~\citep{1969Natur.222..229R}, glitches have been regularly observed and generally studied. One of the earliest proposals for glitches is the starquake model~\citep{BAYM1971816}, in which the gradual buildup of elastic stress inside the star is suddenly released through sudden fracture of the crust, leading to a change in the stellar moment of inertia and consequently a spin-up event.
Motivated by this scenario, we are investigating whether a similar mechanism can work when the dominant driving force is magnetic rather than gravitational, to be applicable to both conventional neutron stars and strangeon stars.

An interesting point of magnetically driven starquakes over conventional gravity-driven ones is the potential to account for  ``{\em anti-glitches}''. In contrasts to glitches, anti-glitches are characterized by sudden decreases in the rotational frequency of a pulsar. The first anti-glitch event was reported in the magnetar 1E2259+586~\citep{Archibald2013}, and subsequent detections in other magnetars and even in rotation-powered pulsars suggest that anti-glitches are not restricted to a single pulsar population \citep{Hu2024,2024ApJ...967L..13T,2025A&A...697A.178P,2025ApJ...991L..18W}. It is easy to argue that gravity-driven starquake models naturally tend to decrease the stellar moment of inertia and therefore can only explain glitches.
Magnetically driven starquakes, however, are under no obligation to decrease the moment, because the internal magnetic field of a magnetar is expected to posses a highly complex configuration and the stresses acting on the star may oppose or even outweigh the effects of gravity in certain regions, for either normal neutron star~\citep[e.g.,][]{1992ApJ...392L...9D} or strange star~\citep{2001A&A...371..963X}.
Accordingly, magnetically induced deformations may either increase or decrease the moment of inertia, making both glitches and anti-glitches both possible within such framework.

While magnetically driven starquakes may naturally account for both glitches and anti-glitches, their viability depends also on whether starquakes can generate rotational-frequency changes that match the observed magnitude, a challenge that is closely related to the internal structure of the pulsar.
In the standard model of neutron stars, only the thin solid crust is capable of sustaining elastic stress. Consequently, only a small fraction of the stellar matter can be involved in a starquake, severly limiting the maximum change in the stellar moment of inertia through elastic deformation.
As a result, conventional gravitational starquake models generally struggle to account for frequent and large glitches, and alternative explanations such as the superfluid vortex model have become the prevailing paradigm.
In contrast, a strangeon star is expected to be globally solid and nearly uniform in density, allowing much more region of the stellar body to contribute stress release.
Starquake in such a compact star may therefore induce substantially larger changes in the moment of inertia~\citep{2004APh....22...73Z}.


In this work, we propose that starquakes, provided their dominant driving stress is magnetic rather than gravitational, can nevertheless give rise to anti-glitches.
We examine whether such model encounters limitations similar to those of conventional gravity-driven starquake models within the neutron-star framework and further investigate whether magnetically driven starquakes in the strangeon-star scenario can generate observable glitch and anti-glitch events.
In Section 2, we introduce our model, which is subsequently applied in Section 3 to account for the observations.
We summarize the conclusions in Section 4.

\section{A toy model}
\label{sect:method}

The relation between magnetic and elastic stresses provides a foundation for modeling a magnetically driven starquake.
It could be derived from either the equilibrium equations of a magnetized pulsar~\citep{2024ApJ...974..125K} or a direct balance between magnetic-stress-induced deformations and elastic restoring stresses~\citep{Lander_2015}. Under suitable approximations, the two approaches lead to equivalent results. In the present work, we adopt the former formulation and derive the stress relation from the equilibrium equations.

Neglecting stellar rotation, we consider the equilibrium stress equations in two scenarios, in cases of both (1) an initial state balanced without elastic and magnetic deformations, and (2) an evolved state balanced with elastic stress because of the magnetic field on a secular timescale. The equilibrium equations of them are, respectively,
\begin{equation}
    \nabla \cdot M^{ij}_{\text{hyd}1} - \rho_1 \nabla \Phi_{\text{G}} = 0, 
\end{equation}
\begin{equation}
    \nabla \cdot (M^{ij}_{\text{hyd}2} + M^{ij}_{\text{elas}} + M^{ij}_{\text{mag}}) - \rho_2 \nabla \Phi_{\text{G}} = 0, 
\end{equation}
where the the stress tensor of perfect hydro-pressure is
\begin{equation}
    M^{ij}_{\text{hyd}1,2} = -p_{1,2}g^{ij}, 
\end{equation}
with $g^{ij}$ the gravitational metric, and $M^{ij}_{\text{elas}}$ ($M^{ij}_{\text{elas}}$) the elastic (magnetic) stress-tensor. We have also ignored the change in the gravitational potential, that is, the Cowling approximation is used.

As demonstrated in previous researches, let
\begin{equation}
    \delta p = p_2 - p_1, \quad
    \delta \rho = \rho_2 - \rho_1, \quad
    \delta M^{ij}_{\text{hyd}} = M^{ij}_{\text{hyd}2} - M^{ij}_{\text{hyd}1},
\end{equation}
one has then an equation for the elastic response to magnetic stress from the difference between above two equilibrium conditions, referring to the work by~\cite{2024ApJ...974..125K},
\begin{equation}
    \label{eq:elastic_response}
    \nabla \cdot (\delta M^{ij}_{\text{hyd}} + M^{ij}_{\text{elas}} + M^{ij}_{\text{mag}}) - \delta \rho \nabla \Phi_{\text{G}} = 0.
\end{equation}
The stress tensors in the formula are given by
\begin{equation}
    \label{eq:exp_elas}
     M^{ij}_{\text{elas}} = -2\mu\sigma^{ij},
\end{equation}
\begin{equation}
    \label{eq:exp_mag}
    M^{ij}_{\text{mag}} = \frac{1}{4\pi} \qty(B^iB^j-\frac{1}{2}g^{ij}B^2),
\end{equation}
where $\mu$ is the shear modulus, with typical value of $3 \times 10^{34}\text{erg/cm}^3$~\citep{Wang_2020}, $\sigma^{ij}$ is the elastic strain tensor, and $B$ is the magnetic field.  

To simplify the problem, we set the pressure and density difference $\delta p,\delta \rho$ to zero, then the equation can be solved easily by setting
\begin{equation}
    \label{eq:mag_elas_equil}
    M^{ij}_{\text{elas}} + M^{ij}_{\text{mag}} = 0.
\end{equation}
Physically, this corresponds to imposing an arbitrary magnetic field on a star that is otherwise in hydrostatic equilibrium without altering the original gravitational balance; the induced magnetic stress is balanced purely by elastic deformation. This approach yields the simplified equation, Eq.(\ref{eq:mag_elas_equil}), governing the elastic displacement induced by magnetic forces.

Next, a criterion for starquake occurrence must be specified. We simply use the von Mises criterion. This criterion could be written as
\begin{equation}
    \label{eq:fracturen_criterion}
    \sqrt{\sigma_{ij}\sigma^{ij}} > \sigma_c.
\end{equation}
Subtituting Eqs. (\ref{eq:exp_elas},\ref{eq:exp_mag},\ref{eq:mag_elas_equil}) to Eq. (\ref{eq:fracturen_criterion}), we obtain
\begin{equation}
    \frac{\sqrt{3}B^2}{16\mu\pi} > \sigma_c,
\end{equation}
where the critical magnetic field that marks the fracture, $B_c$, is
\begin{equation}
    \label{eq:criticalB}
    B_c =\sqrt{\frac{16}{\sqrt{3}}\pi\mu\sigma_c}.
\end{equation}
When the magnetic field strength exceeds this threshold in a region of the stellar crust, a magnetism-driven starquake can be triggered.

\begin{figure}[htbp]
    \centering
    \includegraphics[width=0.5\linewidth]{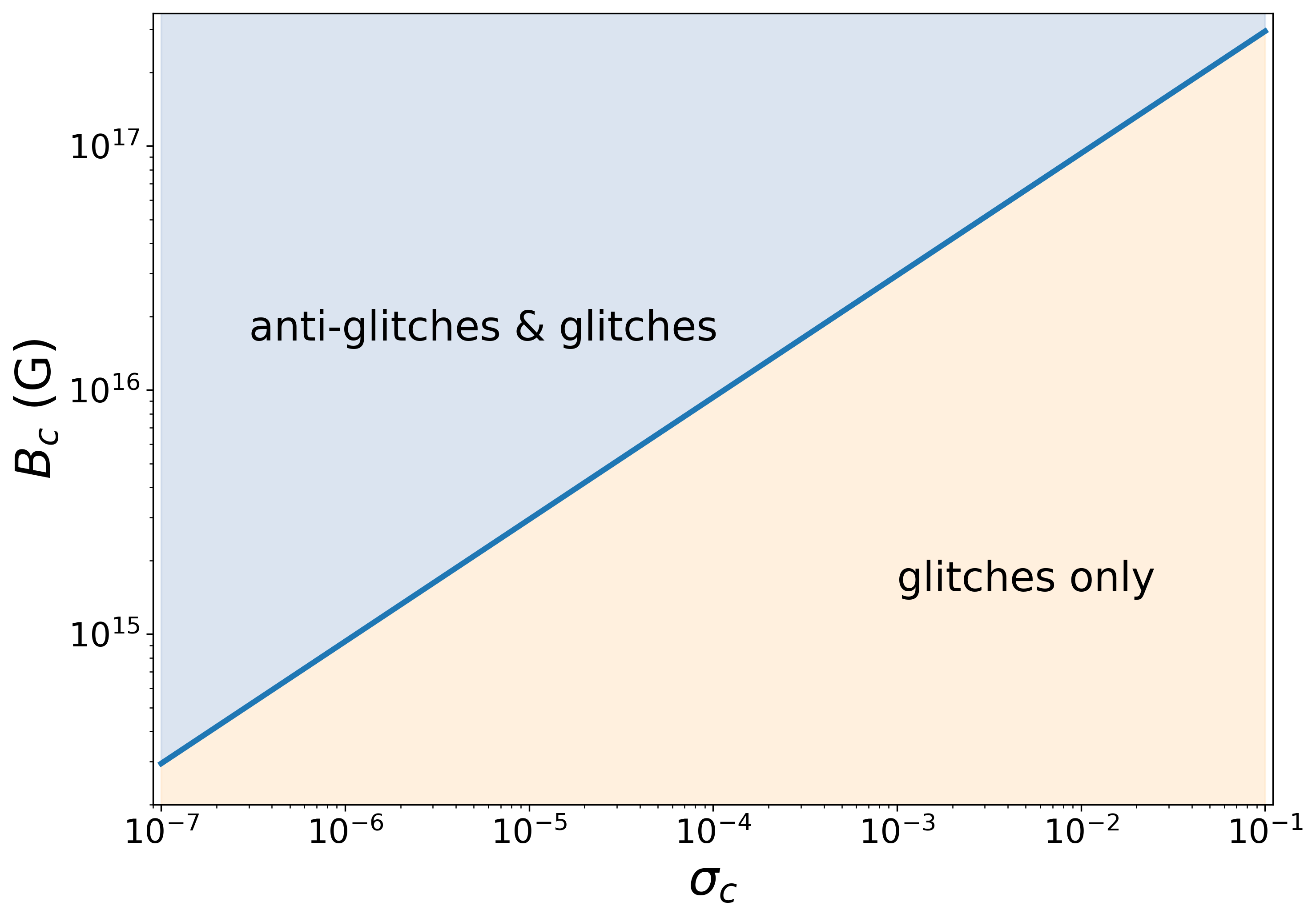}
    \caption{The necessary condition for anti-glitches on $B_c-\sigma_c$ plane. For a certain elastic strain, $\sigma_c$, a magnetism-driven anti-glitch could probably occur if the magnetic field, $B$, is higher than the critical one, $B_c$.}
    \label{fig:Bsigma}
\end{figure}

Molecular-dynamics simulations suggest a characteristic breaking strain of order 0.1, nevertheless, the effective failure threshold in localized regions of the crust may be substantially lower owing to pre-existing defects or remnants of previous fracture events, and the precise value of $\sigma_c$ remains uncertain. Therefore, we treat both $B$ and $\sigma_c$ as parameters, allowing us to draw a curve given by equation(\ref{eq:criticalB}) in the $B$–$\sigma$ plane as shown in Fig.~(\ref{fig:Bsigma}). Above the curve, the magnetism-driven starquakes are expected to occur under our assumptions, thus both anti-glitches and glitches are possible; while below it, they are unlikely, so only glitches are possible. 

We further estimate the relation between $\sigma_c$ and $\Delta\nu/\nu$. We refer to the work by~\cite{Lu_2023} in which the relation is given by
\begin{equation}
\frac{\delta \nu}{\nu} = \frac{\sigma_c \chi^3 R^3 \mu \Gamma}{I},
\end{equation}
where $R$ is the pulsar radius, $\chi$ is defined to be $(r/R)^3$, with $r$ the length scale of the quake region, $\Gamma$ is the ratio of change in moment of inertia to seismic moment with a typical value of $10^{-10}~\mathrm{s}^2$ given in the work for strangeon stars and $10^{-9}~\mathrm{s}^2$ for conventional neutron stars, and $I$ is the moment of inertia of a pulsar. Considering the pulsar as a uniform sphere with mass of a solar mass and radius $R\sim 10$~km, one has $I = 2/5 M_\odot R^2\sim10^{45}\mathrm{g \cdot cm^2}$. Let's consider the ratio, $\xi$, of the glitch amplitude, $\delta \nu/ \nu$, to the crtical strain, $\sigma_c$ (i.e., $\delta \nu / \nu = \xi \sigma_c$), and assume $\chi \approx 0.1$ for a globally solid strangeon star (SS) and $0.01$ for a neutron star (NS) with only solid crust, with typical values of $R \sim 10^6~\mathrm{cm}$, $\mu \sim 10^{34}~\mathrm{erg/cm^3}$, $\Gamma \sim 10^{-10}~\mathrm{s^2}$, and $I\sim10^{45}~\mathrm{g \cdot cm^2}$, we have the ratio
\begin{equation}
\label{eq:sigmac_and_deltanu}
\xi_\text{SS} \approx 10^{-6}, \quad \xi_\text{NS} \approx 10^{-12},
\end{equation}
that implies, compared to a conventional neutron star, a large glitch/anti-glitch for a strangeon star. This results from a higher shear modulus, $\mu$, and larger quake scale, $\chi$, of strangeon stars.

\begin{figure}[htbp]
    \centering
    \includegraphics[width=0.7\linewidth]{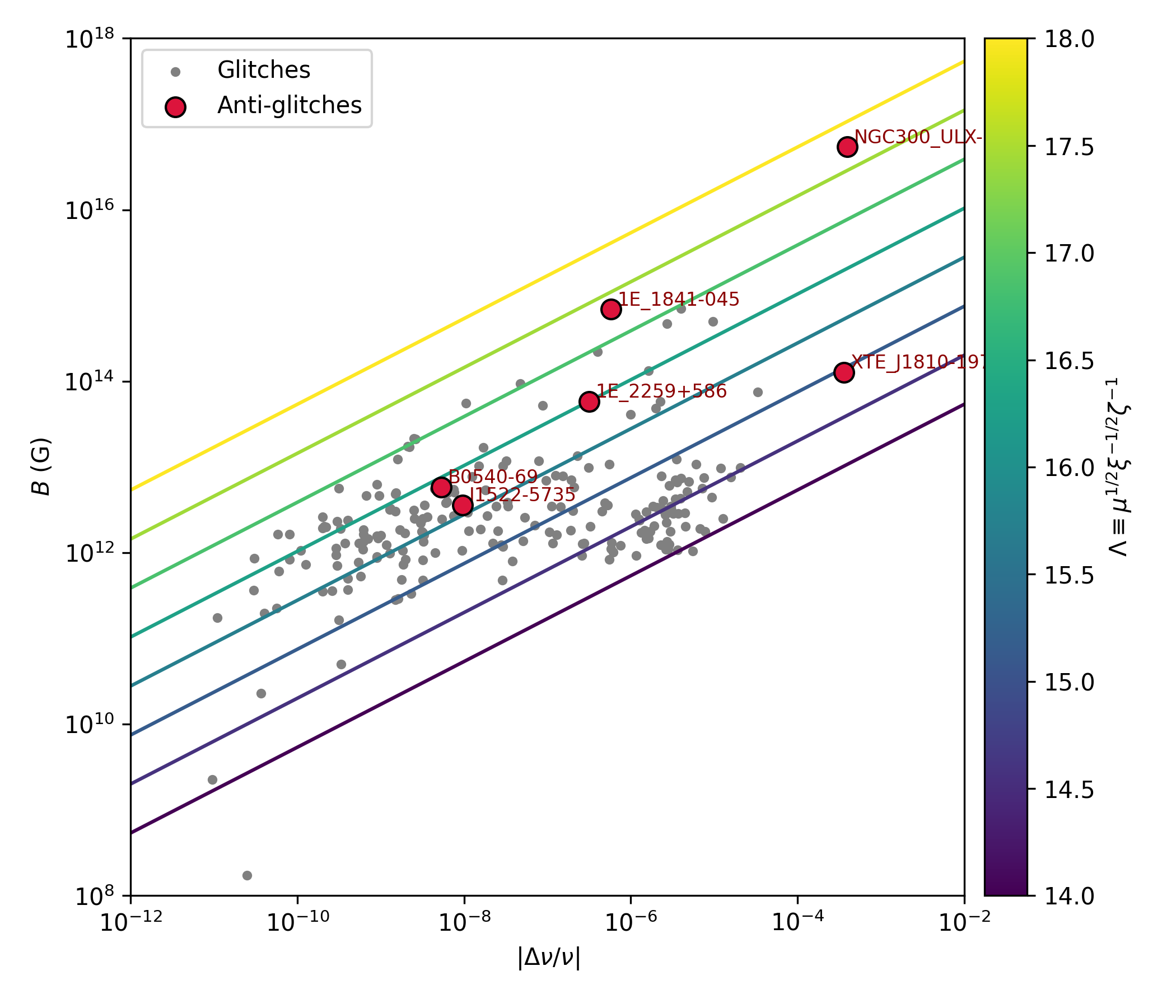}
    \caption{The glitch/anti-glitch condition v.s. observational data.
    For a certain anti-glitch amplitude $|\Delta\nu/\nu|$, the needed $B$-field to make magnetism-driven glitch depends only one parameter, $\Lambda \equiv \zeta^{-1} \xi^{-1/2} \mu^{1/2}$.
    The value of $\Lambda$ ranges from $10^{14}$ to $10^{18}$ (erg/cm$^3$)$^{1/2}$ is applied to fit the observations in the neutron and strangeon star models. Note that PSR B0540-69 and PSR J1522-5735 are not magnetar, and NGC 300 ULX-1 may host a magnetar-strength magnetic field.
    For a pulsars with multiple glitch or anti-glitch events, only the median fractional frequency change ($|\Delta \nu/\nu|$) of all events is shown in the figure.
    The small light-dots represent glitches, while the filled circles for anti-glitch candidates.
    }%
    \label{fig:B_d_sigma_and_data}
\end{figure}

Finally, we can establish an approximate relation between the field $B$ and the amplitude $\Delta\nu/\nu$, which can then be compared to observed glitch and anti-glitch data. Most of the glitches are listed in the ATNF Database~\citep{ATNFGlitchTable}, while the anti-glitch data are collected from the literature: 1E 1841-045, 1E 2259+586, PSR B0540-69, PSR J1522-5735, NGC 300 ULX-1, and XTE J1810-197~\citep{Dib2008_1E1841,Archibald2013AntiGlitch,Ferdman2018B0540,Panin2025J1522,Ray2025NGC300,Camilo2016XTEJ1810}. Note that the $B$-field used in this context refers to the multiple magnetic field in the pulsars, whereas pulsar dipole fields inferred from $P–\dot{P}$ observations may differ by couples of orders of magnitude. Therefore, let‘s introduce a ratio of the multipolar to the dipolar fields, $\zeta = B_\text{m}/B_\text{d}>1$, we can thus construct a $B_d$–$\Delta\nu/\nu$ curve, with $B_d$ the estimated dipole component as shown in Fig.~(\ref{fig:B_d_sigma_and_data}).
It is worth noting that PSR B0540-69 and PSR J1522-5735 are not magnetars, and their anti-glitch behavior may actually be the result of timing irregularities~\citep{2024ApJ...973L..39E}; NGC 300 ULX-1 is generally classified as an ultra-luminous X-ray pulsar rather than a confirmed magnetar~ \citep{Carpano2018,King2019}, but the possibility that it hosts a magnetar-strength magnetic field has been widely discussed by~\cite{Tong2015}. We treat both $\xi$ and $\zeta$ as adjustable parameters, because the numerical estimates of both have an order-of-magnitude error. 

\section{Results and Discussions}
\label{sect:discussion}
Under our hypothesis, anti-glitches are expected to appear above the $B$–$|\Delta\nu/\nu|$ curve in Fig.~(\ref{fig:B_d_sigma_and_data}), which qualitatively aligns with observed data. More importantly, the magnitude of starquakes induced by magnetic stress is expected to increase with magnetic field strength, leading to larger relative changes in rotation rate (i.e., the glitch amplitude $|\Delta\nu/\nu|$), a trend observable in the data. Additionally, some glitches data also exhibit similar behavior.
Nevertheless, it is evident that, the $B$-fields of anti-glitching pulsars are generally much stronger than those of glitching ones.

In Fig.~(\ref{fig:B_d_sigma_and_data}), around \(\Lambda \sim 10^{15}\)~(erg/cm$^3$)$^{1/2}$, our model shows relatively good agreement with the observational data. Some data points exhibit a trend in which the glitch amplitude increases with increasing magnetic-field strength, while others remain nearly unchanged as the magnetic field becomes stronger. Within our theoretical framework, the former corresponds to (anti-)glitches induced by magnetically driven starquakes, whereas the latter corresponds to glitches primarily driven by self-gravity or other factors.\footnote{%
It is worth noting that, pulsars with higher $B$-fields would spin down more quickly, and the stress loading could then be stronger even for quakes  driven by self-gravity.
}%

For conventional neutron stars, adopting \(\xi \sim 10^{-12}\) and \(\mu \sim 10^{30}\,\mathrm{erg\,cm^{-3}}\), one requires \(\zeta \sim 10^{6}\) in order to achieve \(\Lambda \sim 10^{15}\)~(erg/cm$^3$)$^{1/2}$. This implies that the multipolar magnetic field strength must be approximately \(10^{6}\) times stronger than the dipolar field in our model, in order to be consistent with the observational data. In contrast, for strangeon stars, assuming \(\xi \sim 10^{-6}\) and \(\mu \sim 10^{34}\,\mathrm{erg\,cm^{-3}}\), the required condition can already be satisfied when the multipolar magnetic field strength is only about \(10^{2}\) times larger than the dipolar component. In fact, both observational and theoretical studies suggest that magnetars may possess strong multipolar magnetic fields that are typiclly $10\sim 10^2$ times larger than the dipolar components. For example, the low-dipole-field magnetar SGR 0418+5729 has an inferred dipolar field of only \( \sim 6\times10^{12}\,\mathrm{G} \) from spin-down measurements~\citep{Rea2013}, while spectral modeling indicates a surface magnetic field of order \(10^{14}\,\mathrm{G}\)~\citep{Guver2011,Mondal2021}, implying a typical ratio of \( \zeta \sim 10\text{--}10^2 \) which is broadly consistent with the value required in the strangeon-star scenario, but remains far below the ratio of \(\zeta\sim10^6\) required in the conventional neutron stars model.
Therefore, in our approach to understanding anti-glitches, it seems that a conventional neutron star with a crust-fluid-core structure is insufficient to support magnetically driven starquakes capable of producing the required magnitude of observable effects, whereas strangeon stars could naturally satisfy this requirement.

A key difference between starquakes driven by magnetism and gravity is that the former can produce both anti-glitches and glitches, while the latter can only produce glitches.
Quantitatively, magnetically driven events display a positive correlation between $|\Delta\nu/\nu|$ and magnetic field strength, while gravity-driven events show no such dependence.
By grouping the observed glitches and anti-glitches in the $B$–$|\Delta\nu/\nu|$ plot in Fig.~(\ref{fig:B_d_sigma_and_data}), it is possible to identify two clusters: one above the curve, which corresponds to magnetically induced events; and one that is randomly distributed, which corresponds to other mechanisms.
Although the current data does not allow for a definitive conclusion, the general trend shown in Fig.~(\ref{fig:B_d_sigma_and_data}) supports the plausibility of our hypothesis.
Future directions of the research may include: (1) to develop a method of classifying glitches as magnetically driven or other, and then to compare their distribution in the $B$–$|\Delta\nu/\nu|$ diagram; and (2) to refine the qualitative estimates presented in this paper, including the relation between magnetic stress and elastic deformation, the fracture criterion, the link between the fracture limit and the relative change in pulsar spin, and the correspondence between internal multiple and dipole magnetic fields.

Although over a decade has passed since the discovery of the pulsar anti-glitch, the underlying physics remains a subject of heated debate.
Nonetheless, one of the most fundamental issues relevant to this topic concerns the nature of pulsar-like compact objects, namely whether they are conventional neutron stars or strangeon stars (see \cite{2025IJMPA..4050180X} for a recent review).
In addition to the previously discussed methods of identifying strangeon matter, such as mass-radius measurements, constraints from the moment of inertia~\citep{2022MNRAS.509.2758G}, and the detection of strangeon nuggets with acoustic arrays~\citep{2025RAA....25i5010Q}, the research of anti-glitches may provide new insights into this basic study, in which the stellar surface would determine the boundary conditions for pulsar magnetospheres~\citep{2026RAA....26c5014X}, and some peculiar radiative features following an anti-glitch could be observed using highly sensitive telescopes, such the FAST (Five-hundred-meter Aperture Spherical Radio Telescope, \cite{2026ChPhL..43e1103W}).

\section{Conclusions}
\label{sect:conclusion}
We propose a simple and qualitative hypothesis regarding the anti-glitch mechanism: without altering the original equilibrium of a pulsar without a magnetic field, we consider magnetic stresses alone to drive elastic deformation of compact objects with rigidity (either conventional neutron stars with only solid crust, or strangeon stars in a global solid state), and a starquake occurs when the local strain exceeds the breaking threshold, $\sigma_c$, resulting then in an anti-glitch.
In this framework, strong magnetic fields could trigger starquakes, which could increase or decrease the moment of inertia, producing consequently an anti-glitch or a glitch, respectively.
This model makes observable and falsifiable predictions. For example, the magnitude of events should increase with the strength of the internal multipolar magnetic field. Classifying glitch and anti-glitch events according to their correlation with internal $B$-field would provide more direct validation or refutation of this hypothesis in the future.

In summary, this foundational study may address the qualitative potential and scale of magnetically driven starquakes, as well as providing a clear direction for subsequent, more rigorous magnetoelastic numerical investigations.
Last but not least, this approach to explaining the anti-glitch phenomenon may eventually help in  understanding the material world in the Universe~\citep{Xu2024,2025IJMPA..4050180X}: electric matter, strong matter and black holes.

\begin{acknowledgements}
This work is supported by the National SKA Program of China (No. 2020SKA0120100). W.-Y. W. acknowledges support from the NSFC (No.12261141690 and No.12403058), and the Strategic Priority Research Program of the CAS (No. XDB0550300). We would like to acknowledge all the members in the PKU pulsar group for meaningful and enlightening discussions during the the research.

\end{acknowledgements}

\bibliographystyle{raa}
\bibliography{bibtex}

\end{document}